\documentclass[prl,twocolumn,preprintnumbers,amsmath,amssymb]{revtex4}

\usepackage{graphicx}
\usepackage{amsmath,amsfonts,amsthm,amssymb}
\usepackage{physics}
\usepackage{float}
\usepackage{dcolumn}

\def\correspondingauth{\footnote{urbachj@georgetown.edu}}

\begin{document}
\title{Structure of propagating high stress fronts in a shear thickening suspension}

\author{Vikram Rathee}

\affiliation{Department of Physics and Institute for Soft Matter Synthesis and Metrology, Georgetown University,
	Washington, DC 20057 }

\author{Joia M. Miller}
\affiliation{Department of Physics and Institute for Soft Matter Synthesis and Metrology, Georgetown University,
Washington, DC 20057 }

\author{Daniel L. Blair}
\affiliation{Department of Physics and Institute for Soft Matter Synthesis and Metrology, Georgetown University,
Washington, DC 20057 }

\author{Jeffrey S. Urbach \correspondingauth{}}
\affiliation{Department of Physics and Institute for Soft Matter Synthesis and Metrology, Georgetown University,
Washington, DC 20057
}

\begin{abstract}
We report direct measurements of spatially resolved stress at the boundary of a shear thickening cornstarch suspension revealing persistent regions of high local stress propagating in the flow direction at the speed of the top boundary.  The persistence of these propagating fronts enables precise measurements of their structure, including the profile of boundary stress measured by Boundary Stress Microscopy (BSM) and  the non-affine velocity of particles at the bottom boundary of the suspension measured by particle image velocimetry (PIV).  In addition, we directly measure the relative flow between the particle phase and the suspending fluid (fluid migration) and find the migration is highly localized to the fronts and changes direction across the front, indicating that  the fronts are composed of a localized region of high dilatant pressure and low particle concentration.  The magnitude of the flow indicates that the  pore pressure difference driving the fluid migration  is comparable to the critical shear stress for the onset of shear thickening.  The propagating fronts fully account for the increase in viscosity with applied stress reported by the rheometer and are consistent with the existence of a stable jammed region in contact with one boundary of the system that generates a propagating network of percolated frictional contacts  spanning the gap between the rheometer plates and producing strong localized dilatant pressure.
\end{abstract}
\maketitle

\section{Introduction}

Dense suspensions can show complex nonlinear flow behavior such as shear thickening, an increase in viscosity, $\eta$, above a material dependent critical shear stress (reviewed in \cite{Morris:2020aa}). At sufficiently high particle concentration, characterized by the volume fraction of the solid phase, $\phi=V_{\rm{particles}}/V_{\rm{total}}$, the suspension viscosity can increase by more than an order of magnitude for small increases in the applied stress above the critical value. This phenomena represents an interesting example of a nonequilibrium phase transition \cite{Sedes:2020vg} with important implications for practical applications \cite{Morris:2020aa}. 

The average properties of shear thickening suspensions have been extensively studied \cite{Barnes:1989wq,Mewis:2009uu,Morris:2020aa}, and many aspects of the dependence of the transition on the nature of the particulate phase and the particle interactions and on measurement geometry and boundary conditions have been clarified \cite{Morris:2020aa}.  Dramatic shear thickening is now understood to arise from 
 a transition from primarily hydrodynamic interactions at low stress to primarily frictional interactions when 
the stress is high enough to overcome inter-particle repulsion \cite{Wyart:2014wy,Wyart.2014zf,Mari:2015tn,Morris:2020aa}.  Using this framework, many aspects of bulk rheological measurements can be understood, including the complex temporal dynamics observed in measurements of shear thickening suspensions \cite{Richards:2019th}.

A full picture of shear thickening that includes spatial as well as temporal dynamics has remained elusive. Evidence for nontrivial spatiotemporal fluctuations has been found in experiments \cite{Lootens:2003ur,Guy:2015wc} and computer simulations \cite{Heussinger:2013vd,Grob:2016we,Saw:2020va,Rahbari:2021aa}.  Recent experiments using ultrasound imaging of a sheared cornstarch suspension revealed localized velocity fluctuations that form bands which travel along the vorticity direction (perpendicular to both the direction of flow and the shear gradient) \cite{Saint-Michel:2018vw}.   X-ray imaging of cornstarch suspensions showed the existence of concentration fluctuations that appear as periodic waves moving in the flow direction \cite{Ovarlez:2020ul}. Direct observation of the surface of a sheared suspension revealed localized dilation-induced surface deformations \cite{Maharjan:2021ud}.
Using Boundary Stress Microscopy (BSM), a technique that allows for direct measurement of the spatially resolved stresses at the surface of a sheared suspension, we have shown that the increase in stress associated with shear thickening in suspensions of silica particles is the result of localized transitions to high viscosity or fully jammed phases \cite{Rathee:2017aa,Rathee:2020aa}.

Connecting these observations to quantitative models that have been developed to describe average behavior will require continuum descriptions that can move beyond mean field and include non-affine flow, concentration fluctuations, and the coupling between flow, concentration, and stress.  Here we describe a regime of shear thickening in cornstarch suspensions that displays persistent  regions of high local stress that propagate in the flow direction.  The persistence of these regions enables a detailed description of their structure, including the profile of boundary stress, the non-affine velocity of particles at the boundary where local stress is measured, and a direct measurement of the relative flow between the particle phase and the suspending fluid.  These observations provide a striking example of the complex structures associated with shear thickening and should help guide the development of models that can capture that complexity.

\section{Shear thickening via propagating fronts}

\begin{figure*}
\includegraphics[width=0.8\textwidth]{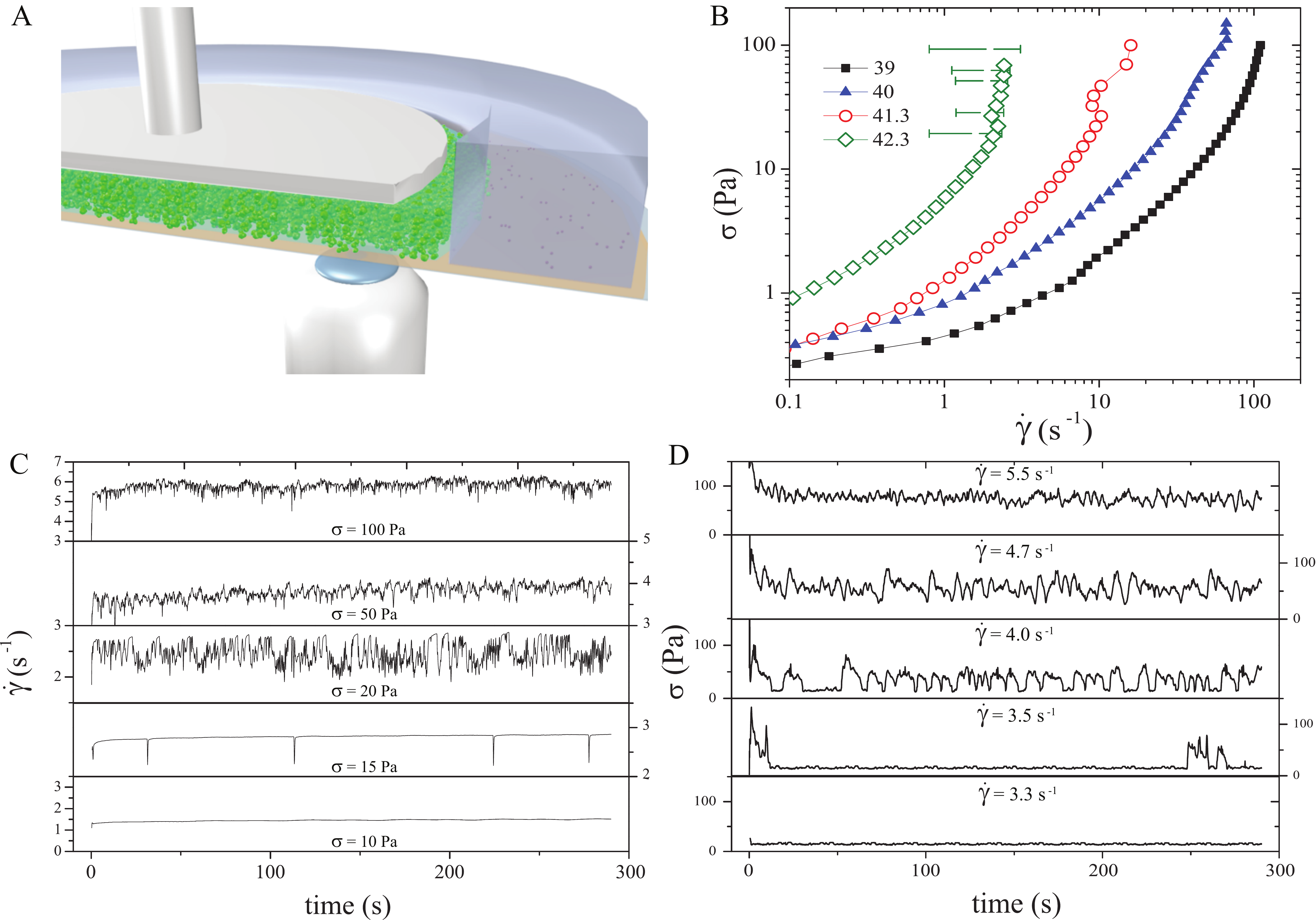}
\caption{Rheology of cornstarch suspensions.  (A) Schematic of setup for performing simultaneous rheology, particle velocimetry, and boundary stress microscopy (BSM). (B) Flow curves (stress vs strain rate $\dot\gamma$) at different weight fractions (\% indicated in legend).
The error bars for $\phi _w= 42.3$\% indicate the range of shear rates measured when the stress is held constant.  (C) Shear rate vs. time at constant applied stress ($\phi _w= 42.3$\%). (D) Stress vs. time at constant shear rate ($\phi _w= 42.3$\%).}
\label{bulkrheology}
\end{figure*}

We investigate a well-studied model system, cornstarch grains suspended in a density matched glycerol-water-cesium chloride mixture (see Methods for details).  The cornstarch is highly polydisperse, with a mean grain diameter of about 20 $\mu$m. The suspension is subjected to shear with a parallel plate rheometer, with modifications to allow for simultaneous imaging (Fig \ref{bulkrheology}A, \cite{Dutta:2013aa}).  Note that the sample extends beyond the outer edge of the rheometer tool, which does have an impact on  bulk rheology measurements \cite{Fall:2012vd}, as discussed below.  A typical set of flow curves, showing the rheometer stress $\sigma$ versus average shear rate $\dot\gamma$ for range of applied stresses at different concentrations,  is shown in Fig. \ref{bulkrheology}B.  Figure \ref{bulkrheology}C shows $\dot\gamma$ as a function of time for the sample at a weight fraction of $\phi _w = 42.3\%$ (corresponding to a volume fraction of $\phi \approx 0.48$), where the critical stress for shear thickening $\sigma _c \approx 15$ Pa.  Similar to previous results \cite{Hermes:2016wj}, shear rate fluctuations increase dramatically in the shear thickening regime. Fig. \ref{bulkrheology}D shows the stress as a function of time for the same sample as in (c), but with shear rate held constant.  Again we see large fluctuations above a critical shear rate, which can be understood as the shear rate necessary to produce stresses above $\sigma _c$. These rheology measurements are consistent with previous reports \cite{Fall:2012vd,Richards:2019th}, with minor quantitative differences that can be attributed to differences in the source material, measurement configuration, and boundary conditions.  

\begin{figure*}
\includegraphics[width=1\textwidth]{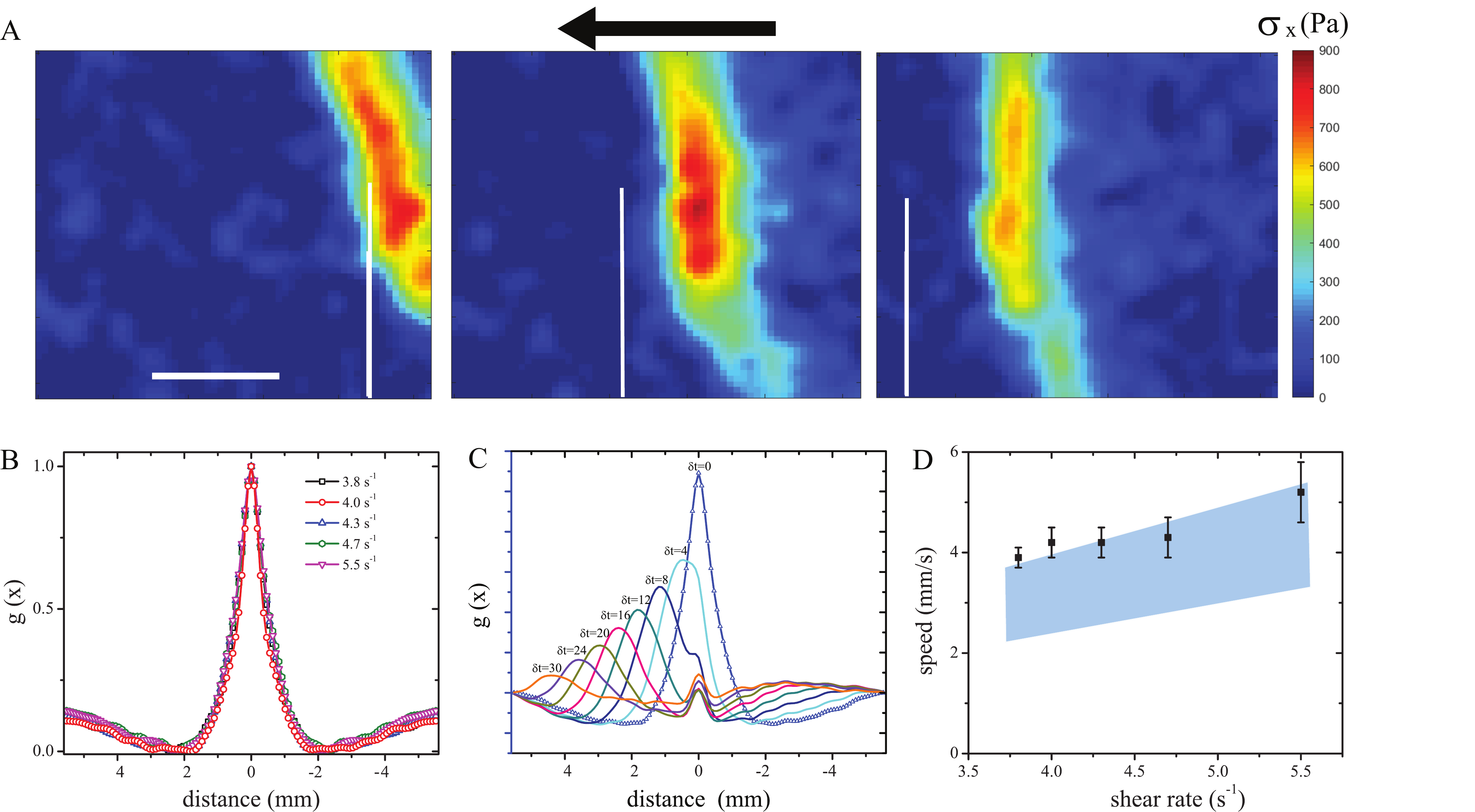}
\caption{Fronts of high stress propagate at the speed of the rheometer tool.   (a) Spatial maps of boundary stress component in the flow direction ($\sigma _x$) at three consecutive timepoints with a constant applied shear rate $\dot\gamma =4 s{-1}$.  (Scale bar 2 mm, arrow indicates the direction of flow). (b) Average spatial autocorrelation of $\sigma _x$ measured in a horizontal strip in the center of the boundary stress field at different values of $\dot\gamma$. (c)  Spatio-temoral cross-correlation of $\sigma _x$ at $\dot\gamma =4 s{-1}$ a for different delay times (in number of frames, 36 ms/frame).  (d) Propagation speeds extracted from peaks in cross-correlation curves as a function of $\dot\gamma$. The shaded band represents the range of speeds  of rheometer top plate in the central region of stress maps. $\phi =42.3\%$.}
\label{lowmag} 
\end{figure*}

Spatial heterogeneities underlie the dramatic increase in stress, as can be revealed by Boundary Stress Microscopy (BSM, \cite{Arevalo:2015aa}). To perform BSM,  the usual bottom rheometer plate is replaced with a glass slide coated with a thin, uniform transparent elastomer (PDMS) of known elastic modulus with a sparse coating of chemically bound fluorescent microspheres. The measured displacements of the microspheres are used to determine the stresses at the boundary with high spatial and temporal resolution (Methods).  When the rheometer stress is below $\sigma _c$, the boundary stress is uniform and low, similar to previous measurements of colloidal silica spheres \cite{Rathee:2017aa,Rathee:2020aa,Rathee:2020un}.  In contrast, above $\sigma _c$ we observe regions of high boundary stress that extend in the vorticity direction and propagate in the flow direction.  Figure \ref{lowmag}A shows the map of boundary stress at three successive time points (see also Supplemental Movie 1) for an applied shear rate of $\dot\gamma = 4 s^{-1}$.  The map covers an area of 5.7 mm$^2$ on the outer half of the 12.5 mm radius rheometer tool.  The peak stresses are about an order of magnitude higher than the average stress, similar to previous measurements of colloidal silica spheres \cite{Rathee:2017aa,Rathee:2020aa,Rathee:2020un}.

High stress fronts appear above $\sigma _c$ for both constant applied shear rate (Supplemental Movie 1 \cite{SM_movie_link}) and constant applied stress, though their propagation speed and spatial extent are more stable at constant shear rate. We quantify their average behavior as a function of applied shear rate by calculating the spatio-temporal cross-correlation of the boundary stress fields measured for a constant shear rate applied for roughly 290 seconds.   Figure \ref{lowmag}B shows the spatial autocorrelation of $\sigma_x (x)$ calculated for a horizontal strip in the middle (vertically) of the image.  The spatial stress profile is remarkably independent of shear rate, and  the spatial extent is roughly equal to the gap between the rheometer plates, consistent with previous measurement on silica sphere suspensions \cite{Rathee:2017aa,Rathee:2020aa,Rathee:2020un}. Measurements of the spatio-temporal cross-correlation at different time lags, $\sigma_x (\delta x,\delta t)$ (Fig. \ref{lowmag}C) reveal  the average propagation in the flow direction, from which we can extract the propagation speed as a function of applied shear rate, as shown in Fig. \ref{lowmag}D.  The shaded region shows the speed of the top plate in the imaged region, which closely matches the speed of the propagating fronts. (The speed of the outer radius of the rheometer tool is $\dot\gamma h$, where the gap  $h$ between the rheometer plates is 1 mm. The outer edge of the imaged region, at the bottom of the figures, is about 1 mm inside  of the outer radius.) The implications of the steady propagation speed and the connection with prior measurements are discussed below.

\begin{figure}
	\includegraphics[width=\columnwidth]{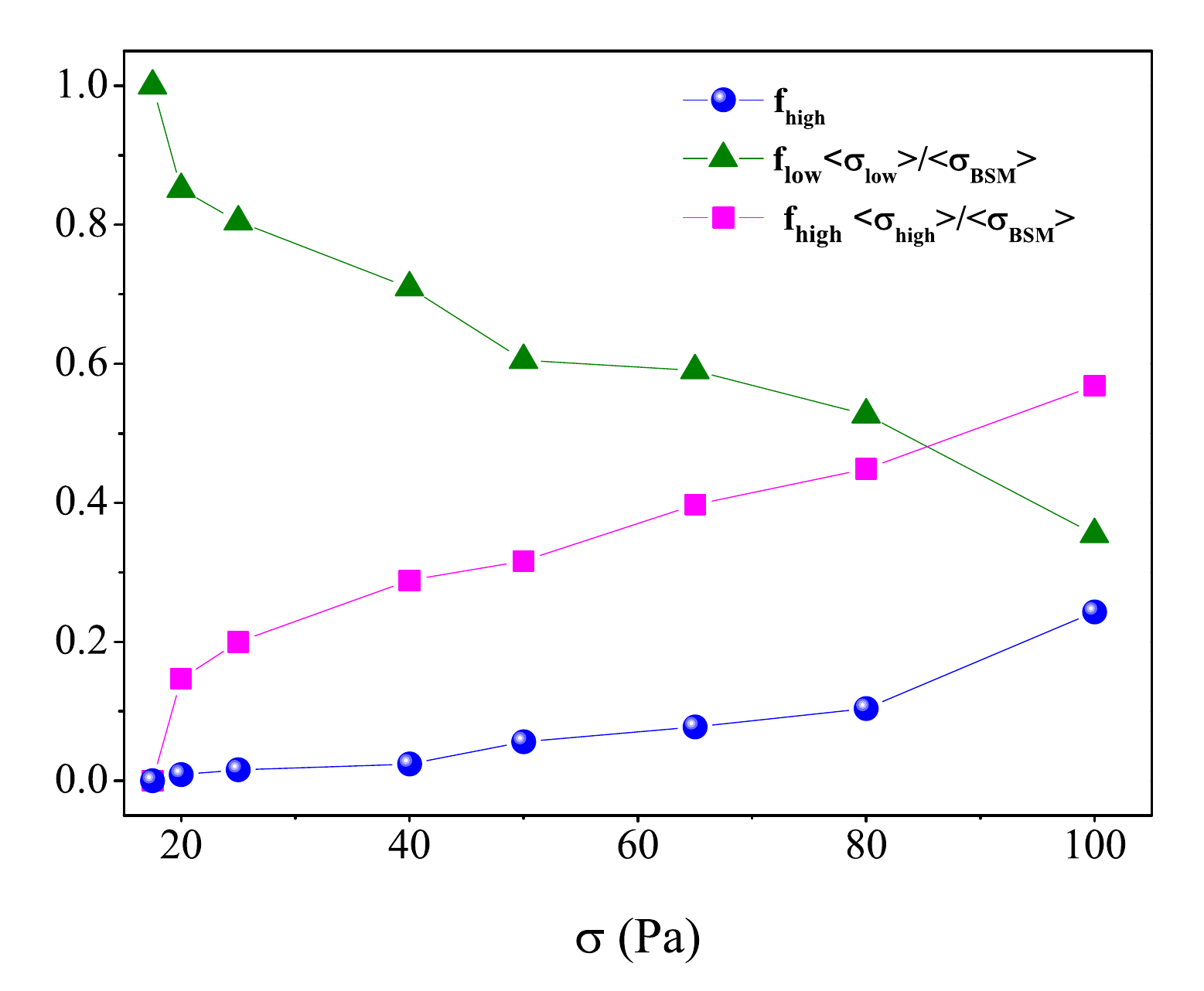}
\caption{High stress fronts are responsible for shear thickening.  The  area fraction of the stress maps containing high stresses (blue circles) is negligible at applied stresses at or below $\sigma _c$, but rises rapidly with applied stress, reaching over 20\% of the area at 100 Pa. The stress contribution from the high stress regions (magenta squares) grows as the applied stress is increased at the expense of the contribution from the low stress regions (green triangles) until it becomes the dominant source of stress. } 
\label{segmented}
\end{figure}

The high stress regions contribute to the shear thickening by becoming more stable and more frequent, in a manner most clearly observed from measurements at constant applied stress.  Well above $\sigma _c$, the fronts of high stress propagate like  those observed under constant applied shear rate. At stresses closer to $\sigma _c$, the regions are both less frequent and less stable and form, evolve, and disappear within the field of view.  The large difference in boundary stress between the high stress fronts and the background allows a clear identification of the contribution of the fronts to shear thickening using image segmentation (Methods).  The average stress within the fronts is roughly constant as a function of applied stress, while the background stress remains small.  A plot of the fraction of the boundary covered by the high stress regions vs. stress shows the dramatic increase in area fraction as the applied stress is increased above $\sigma_c$ (Figure \ref{segmented}, blue circles).  The fractional contribution of the two regions to the total stress measured by BSM is also plotted in Fig. \ref{segmented}, and shows that the proliferation of localized high stress regions accounts for the rapid rise in suspension viscosity, similar to our previous measurements of suspensions of silica spheres \cite{Rathee:2017aa}.

Finally, we note that while stable propagating fronts are a robust phenomena in the parameter range reported here, we have observed more complex behavior in other conditions.  In particular, for rheometer gaps below $\sim 0.5$ mm, we see high stress regions migrating in the vorticity direction as well as regions propagating in the flow direction.  At larger gaps ($>1.5$ mm) and shear rates well above critical value ($>6$ s$^{-1}$) the propagating front motion appears to be more variable.  As discussed below, this is consistent with the general picture that emerges from a variety of sources that the spatiotemporal dynamics of the transitions underlying shear thickening are sensitive to the measurement geometry.

\section{Profiles of boundary stress and boundary velocity}
\label{profiles}
Using the combination of rheology with high speed confocal microscopy \cite{Dutta:2013aa} we are able to directly investigate the dynamics of the high stress regions.   Fluorescein  sodium salt is added to the solvent is absorbed by the porous cornstarch particles, providing a clear image of the bottom layer of the suspension (Supplemental Movie 2 \cite{SM_movie_link}).  The suspension is not index matched so we cannot image beyond the bottom layer of particles.

\begin{figure}
\includegraphics[width=1\columnwidth]{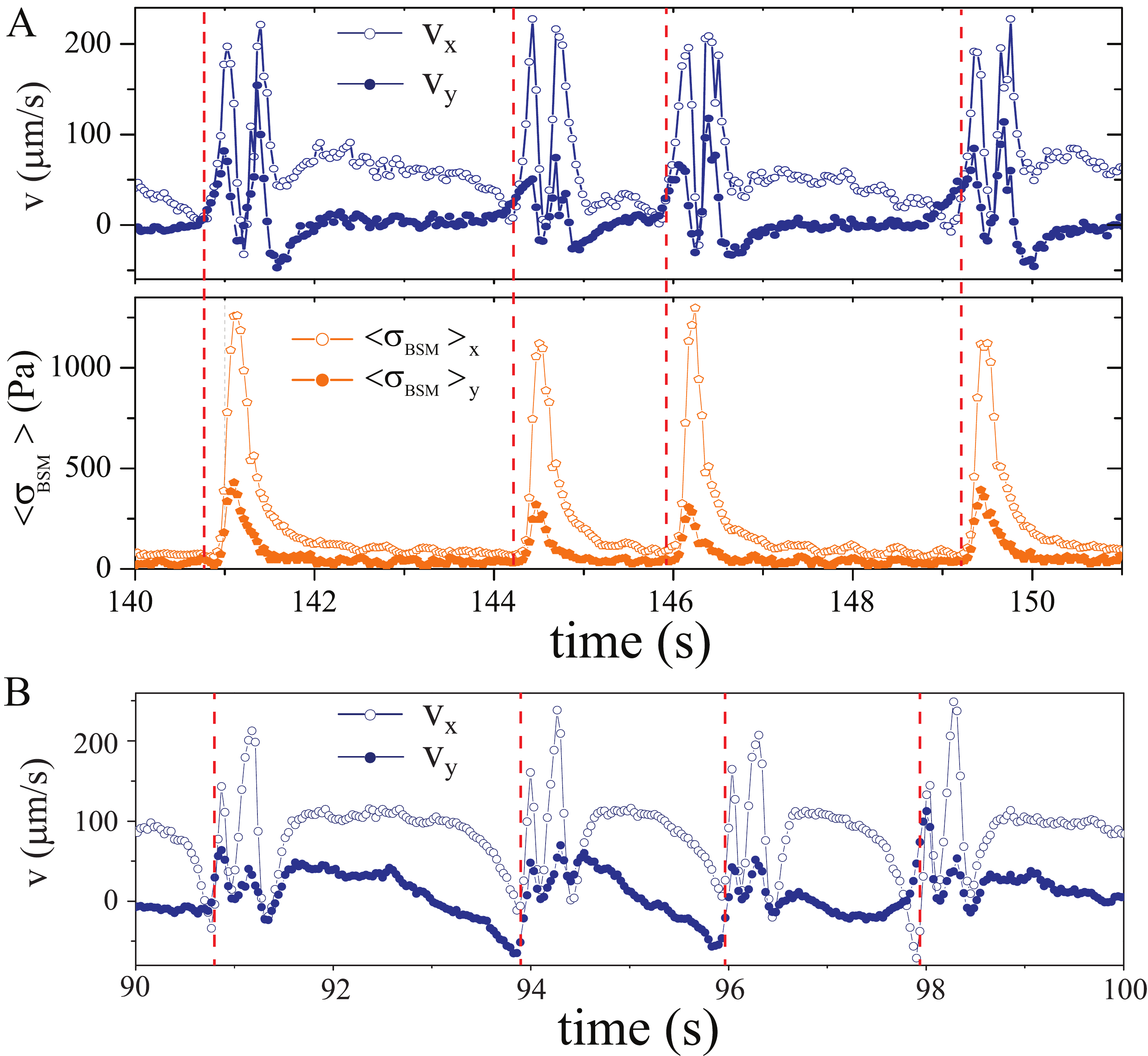}
\caption{Nonaffine velocity fluctuations of cornstarch bottom layer.   A) Top graph: Average of the $x$ (flow, open circles) and $y$ (vorticity, filled circles) components of the cornstarch velocity determined by PIV of successive images.   Bottom graph: Components of the boundary stress in the $x$ (open circles) and $y$ (closed circles) directions.  Dashed lines indicate the time points identified as $t=0$ for the high stress event averaging used to generate Fig. \ref{average}.  (PDMS substrate, 100 Pa applied stress). B) Average of the $x$ and $y$ components cornstarch velocity measured on a plain glass substrate (100 Pa applied stress). Dashed lines indicate the time points identified as $t=0$ for the high stress event averaging used to generate Fig. \ref{average}.}
\label{velocity}
\end{figure}

For stresses below $\sigma _c$, the cornstarch moves uniformly in the flow direction with an approximately constant velocity comparable to $R \dot\gamma$, where $R$ is the average particle radius of 10 $\mu$m and $\dot\gamma$ is the average shear rate reported by the rheometer, consistent with affine flow for the suspension. When the applied stress exceeds $\sigma _c$, the affine flow is interrupted by short periods where the flow changes dramatically.  The top panel of Figure \ref{velocity}A shows part of a typical time series of the velocity determined by PIV containing 4 of these events (see also Supplemental Movie 2 \cite{SM_movie_link}).  The velocity in the flow direction ($v_x$)  shows a distinctive M-shape, with two sharp peaks in speed. These events are accompanied by substantial variations of the velocity component in the vorticity direction ($v_y$), which shows a similar M-shape, but with a smaller magnitude.  The positive values of $v_y$ represent flow towards to outer boundary of the rheometer, suggesting that the suspension is bulging out due to dilatant pressure, as discussed below.   The bottom pane of Fig. \ref{velocity}A  shows the stresses measured by BSM performed simultaneously with the PIV, where each data point represents an average of the stress component in each image. While the area of these high magnification images is small compared to the images show in Fig. \ref{lowmag}, the boundary stress peaks associated with high stress fronts are still easily measurable.  The velocity fluctuations are always coincident with high boundary stress.  In order to confirm that this behavior is not sensitive to the compliance of the substrate, we have performed the same measurements on an elastic layer 5 times stiffer and found essentially identical behavior (Supplemental Figure 1).  In addition, we have carried out  measurements on uncoated glass (without PDMS), and again found nearly identical features in the velocity time series (Fig. \ref{velocity}B, Movie 2, 2nd half \cite{SM_movie_link}). 

\begin{figure}
\includegraphics[width=1\columnwidth]{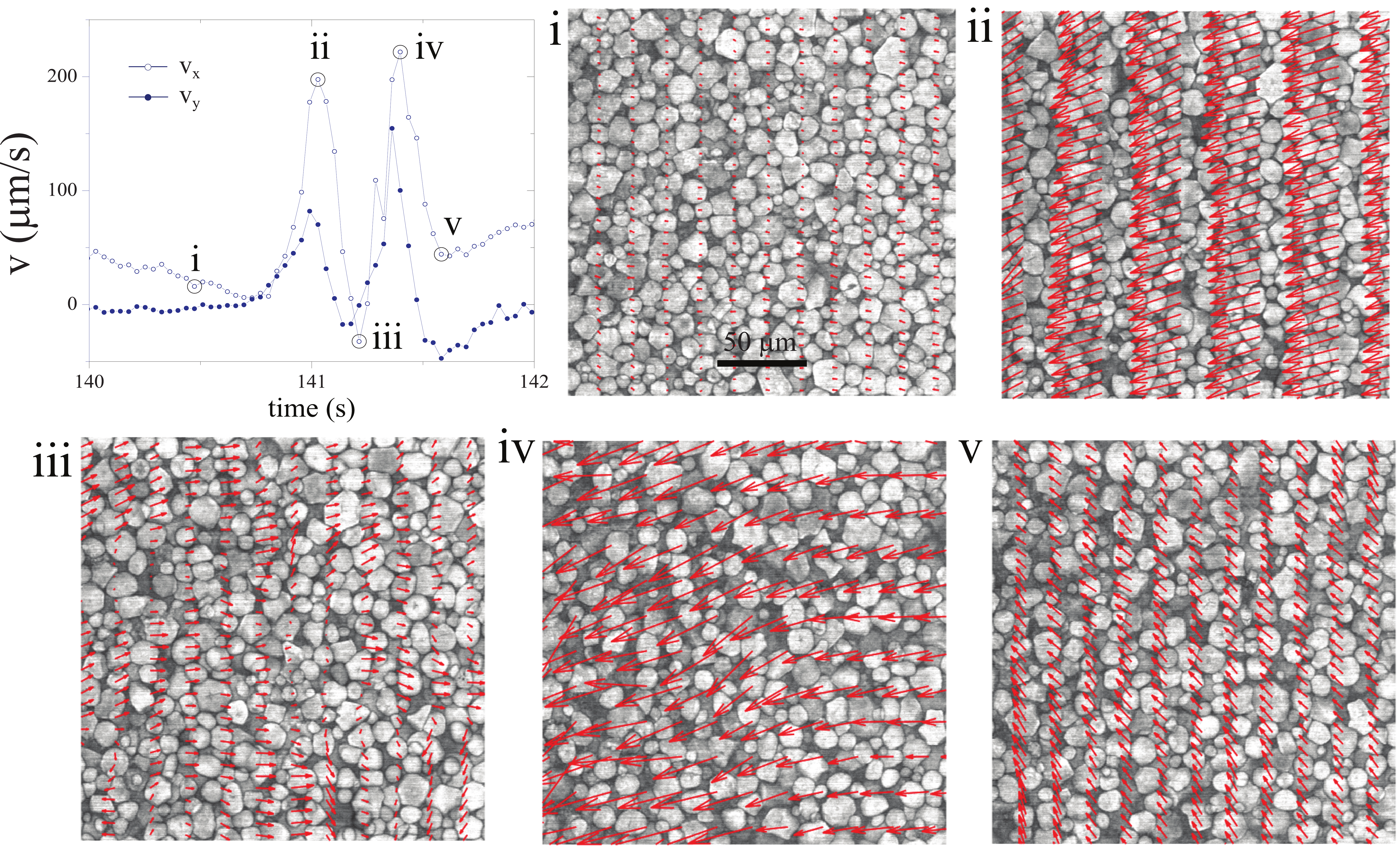}
\caption{Flow fields during the first event in Fig.\ref{velocity}A.  The images i-v show the cornstarch particles at the time points circled on the plot of the average $v_x$ shown in the first panel, with superimposed displacements calculated by PIV.}
\label{pivimages}
\end{figure}

The flow fields measured during a typical event are displayed in Fig.  \ref{pivimages}.  The cornstarch flow speed gradually slows while remaining mostly affine (i) before sharply increasing in both the flow and vorticity directions (ii), then slowing almost to a standstill (iii), speeding up again (iv), and finally relaxing to almost affine flow (v).

The variations in boundary layer velocity and boundary stress are associated with more subtle variations that are evident on visual inspection (Supplemental Movie 2 \cite{SM_movie_link}).  Preceding the events, the layer of particles gradually becomes more dense, until there is very little space between the cornstarch particles.  The particles mostly move together during the events, but there is a point where they separate and rotate and then settle in to a less dense layer which then starts to compactify as the  the next event approaches.  Finally, by watching the motion of the fiduciary beads attached to the elastic layer (Movie 3 \cite{SM_movie_link}), it is clear that the middle of M-shape represents a region where the cornstarch particles briefly come to rest relative to the PDMS, while both the PDMS and the cornstarch particles move relative the laboratory reference frame, indicating direct physical contact with no intervening fluid layer.  Interestingly, the compliance of the boundary does not significantly impact the velocity profile, as nearly identical profiles are observed on glass substrates (Fig. \ref{velocity} and Movie 2 \cite{SM_movie_link}).  

\begin{figure*}
\includegraphics[width=1\textwidth]{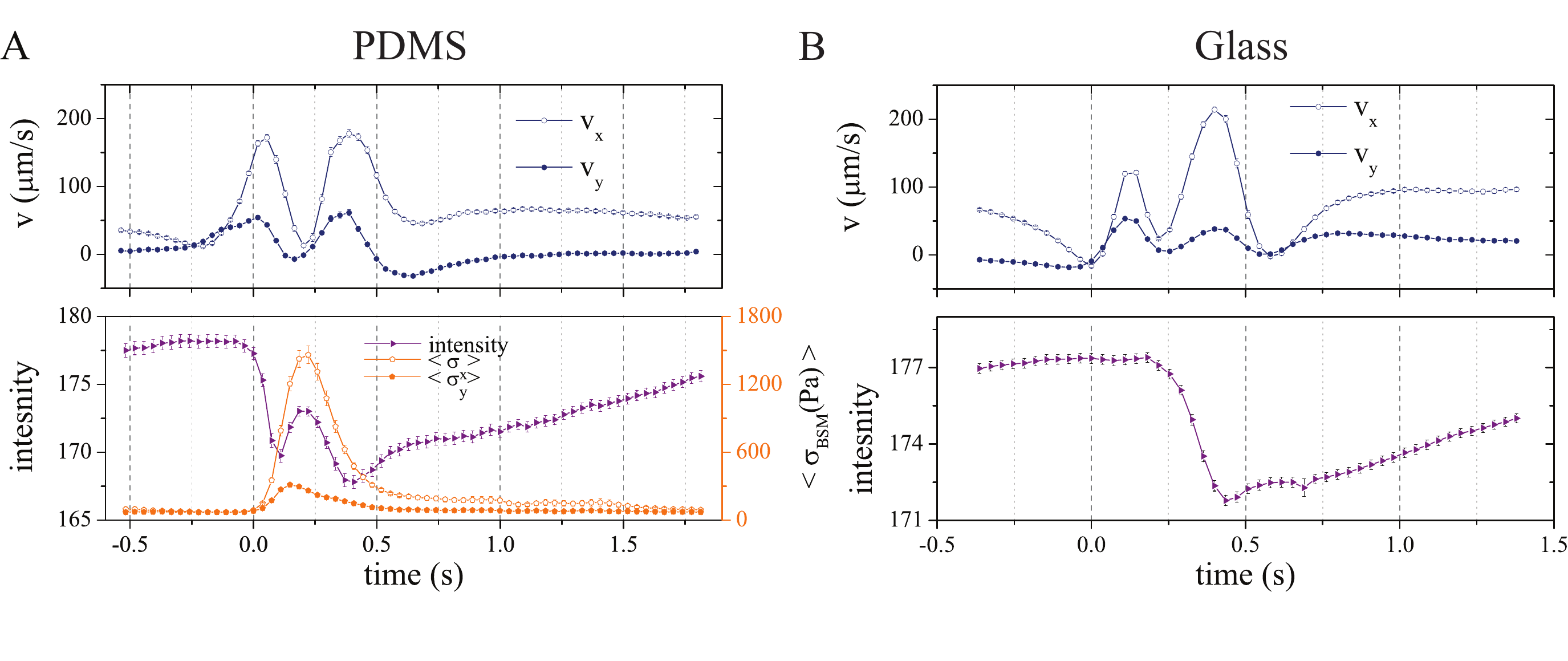}
\caption{Average event shape for (A) PDMS coated bottom surface and (B) uncoated glass.  $v_{x,y}$ represent the components of the velocity of the bottom layer of conrnstarch particles in the flow and vorticity directions, respectively, measured by PIV (as in Fig. \ref{velocity}), while $<\sigma_{x,y}>$ shows the image-averaged components of the boundary stress.  The intensity values, calculated directily from the images of the cornstarch particles, provide a measure of the cornstarch concentration in the bottom layer. 
} 
\label{average}
\end{figure*}

The high degree of regularity of the features associated with the high stress events enables a precise characterization of their properties.  As detailed in Methods, we identify the onset of an event as a rapid increase in boundary stress from a low background (dashed lines in Fig. \ref{velocity}B).  For measurements on a glass substrate, the first minimum of the M-shape provides a robust identifier (dashed lines in Fig. \ref{velocity}C). 
Using these algorithmically identified time points, we can calculate the characteristic event shapes by averaging  the  boundary stress and cornstarch velocity measured at times relative to those points. 
Figure \ref{average}A  shows the average of the cornstarch velocity fields for all events detected during 290 seconds at 100 Pa applied stress (N=80), demonstrating the M-shape described above, as well as revealing a modest but clear decrease in speed in the flow direction approaching the event, and an inward flow (in the negative $y$ direction) following the event. Figure \ref{average}B shows a similar analysis for the PIV time series on a smooth glass substrate, where the shape of the events closely matches those observed on the elastic substrate.
The bottom panel of Fig. \ref{average}A
shows the result for the $x$ and $y$ components of the average the boundary stress for all of the events on the elastic substrate. The component of the stress in the flow ($x$) direction shows a rapid increase followed by a slow decay.  The stress in the vorticity ($y$) direction is smaller but has a similar shape, and represents the suspension pulling the boundary in the direction of the outer edge of the rheometer.   The bottom panel of Fig. \ref{average}A also shows the average intensity of the cornstarch image, which is roughly proportional to the concentration of particles in the bottom layer.  The intensity increases approaching the event, reflecting the compacting of the bottom layer, followed by a rapid change during the event that ends with a minimum in the intensity, followed by a slow increase as the layer compacts again.   The bottom panel of Figure \ref{average}B shows  the average intensity profile measured on the glass substrate, which reveals a similar compaction but a simpler overall shape.

One subtle but clear feature of the velocity profiles is a gradual reduction in $v_x$ approaching the events. The videos suggest that the grains on the bottom layer mostly orient such that they have a face parallel to the bottom boundary, creating a roughly uniform layer of cornstarch a small distance above the bottom surface.  The thickness $d$ of the layer and $v_x$ determine the local stress on the bottom surface according to $\sigma=\eta_s v_s/d$, where $\eta _s$ is the solvent viscosity. As discussed above, away from the events we expect that the boundary stress is comparable to the critical stress $\sigma _c\approx 20$ Pa, which implies fluid layer with thickness $d\sim 50$ nm. We do not observe a significant decrease in local stress approaching the events (although the small stresses are close to our measurement resolution), so one interpretation of the reduction in $v_x$ is that there is a modest increase in dilatant pressure near the event which generate a reduction in the thickness of the slip layer.  The role of dilatancy is discussed further below. 

\section{Direct measurement of fluid migration}

\begin{figure*}
\includegraphics[width=1\textwidth]{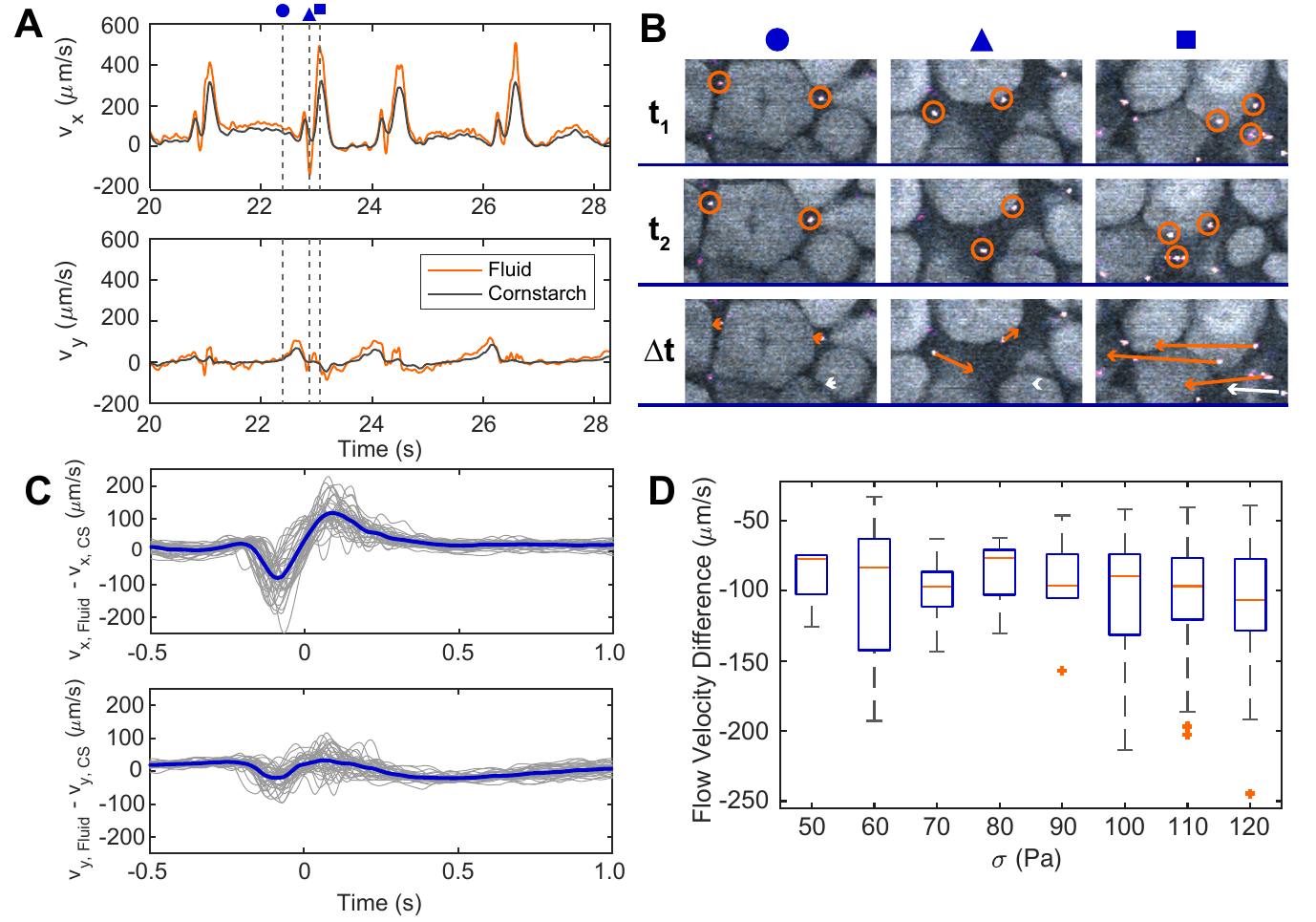}
\caption{
 (A) Time series of cornstarch (grey) and solvent (orange) velocity components in the flow (top) and vorticity (bottom) directions at 120 Pa applies stress. 
(B) Images of cornstarch and tracer microspheres (circled) from high speed video at successive time points ($t_1$ and $t_2$) away from an event (circle, also in A), at the flow velocity minimum (triangle), and at the maximum (square). Bottom row: Cornstarch image at $t_1$ with superimposed arrows at the microsphere positions showing their displacements between $t_1$ and $t_2$. White arrows in the lower right corner of the images show the average cornstarch displacement in the same time interval.
C)  The difference between the cornstarch and microsphere velocities for 36 events, with the thick line representing the average deviation at each time point, where t = 0 is the midpoint between maximum negative and positive differences in flow speed.  
D) Box and whisker plot showing the distribution of the maximum negative value of $v_{x,Fluid}-v_{x,CS}$ as a function of applied stress (63 seconds at each constant applied stress).
} 
\label{solvent}
\end{figure*}

 Since both the suspending liquid and the particulate phases are effectively incompressible, the apparent changes in concentration associated with the cornstarch images should be accompanied by a counter-balancing flow of the solvent.  In order to measure the solvent flow directly,  fluorescent microspheres were suspended in the fluid phase (Methods). Resolving the complex motion of the microspheres during the events requires high imaging speed, which we achieved by reducing the scan range of the laser scanning confocal.  The resulting image sequence (Supplemental Movie 4 \cite{SM_movie_link}) shows that, away from the events, the  fluid and particulate phases move together, with no detectable relative motion. However, during each event we observe a brief period where the solvent flow differs dramatically from the cornstarch.  The  motion of the microspheres during these periods is complex, reflecting the fact that the paths through the cornstarch grains are tortuous, but with high speed imaging we can reliably track particles with standard techniques (Methods).  Figure \ref{solvent}A shows a part of a time series of the average fluid velocity calculated from the microsphere displacements along with the cornstarch velocity measured by PIV, for a constant applied stress of 100 Pa.  Both the $x$ and $y$ components of the two velocities are equal to within the experimental resolution except for a brief period in the middle of the M-shape, when the fluid velocity is less than the cornstarch, sometimes taking on negative $v_x$ values, followed immediately by a period when the fluid velocity is higher than the cornstarch.  To visualize this behavior, Fig. \ref{solvent}B shows successive images ($t_1$ and $t_2$) of the cornstarch with fluorescent microspheres at a time away from an event (circle), during the the minimum of the solvent $v_x$ (triangle) and at its peak value (square).  The final row of images in Fig. \ref{solvent}B shows the image at $t_1$ with superimposed arrows showing the displacement of the microspheres and of the cornstarch between $t_1$ and $t_2$.  Recall that the applied shear flow is from right to left in the images, so the arrows in the second column (triangle) indicate that solvent is moving against the applied flow.
In order to highlight the relative flow between the particulate and fluid phases, the difference between the two velocities is shown in Fig. \ref{solvent}C for 36 events, with the thick line representing the average deviation at each time point.    While there is considerable inter-event variability, the presence of fluid migration in the  $-x$ direction, followed quickly by migration in the $+x$ direction, is a consistent feature of the events.  Similar behavior is observed at applied stresses throughout  the shear thickening regime, with relatively little variation in intensity as measured by the maximum negative value of $v_{x,\rm{Fluid}}-v_{x,\rm{CS}}$ (Fig. \ref{solvent}D).

\section{Discussion}
\label{discussion}

Below we discuss the relationship between the results reported here and recent literature on heterogeneities in shear thickening suspensions, followed by an analysis of the structure of the fronts inferred from our measurements, culminating in a model for fronts that is consistent with that structure.



\textbf{Vorticity flow and dilatancy:} 
The observation that the propagating stress fronts are associated with concentration fluctuations and flow in the vorticity direction is consistent with prior results showing that shear thickening is associated with fluctuations and localized dilatantcy \cite{Boersma:1991wi,Fall:2012vd, Nakanishi:2012ux, Larsen:2014vv, Nagahiro:2013ti, Nagahiro:2016wu, Hermes:2016wj, Richards:2019th, Ovarlez:2020ul, Maharjan:2021ud}.  Fall et al. \cite{,Fall:2012vd} concluded that the excess suspension in Couette  or an over-filled parallel plate rheometer geometries affected bulk rheology by providing a reservoir for particles forced out of the shearing region by dilatant pressure.  Experiments in a Couette geometry and associated continuum modeling have shown that dilatant fronts that propagate in the flow direction can occur in shear thickening \cite{Nakanishi:2012ux,Nagahiro:2016wu}.    Hermes et al. \cite{Hermes:2016wj} observed that rapid decreases in shear
rate were accompanied by local deformations of the air-sample
interface at the edge of the rheometer tool, and that the deformations
sometimes appear static and sometimes move opposite to the direction of
the flow.  Using ultrasound to image velocity and concentration fluctuations during shear thickening in a Couette geometry, Saint-Michel et al.  \cite{Saint-Michel:2018vw} found bands of non-affine velocity and concentration fluctuations propagating in the vorticity direction that  proliferate with increasing stress.
Maharjan et al. \cite{Maharjan:2021ud} report the dynamics of surface roughening in sheared cornstarch suspensions and show that such events underlie the increase in normal and shear stress of discontinuous shear thickening.  Of particular relevance to this study are the results from x-ray imaging of shear thickening suspensions reported in \cite{Ovarlez:2020ul}, where they find concentration fluctuations that appear as periodic waves moving in the direction of flow and are associated with dilatancy. Finally, using a novel approach for directly measuring local normal stresses in sheared cornstarch, Gauthier et al.  \cite{Gauthier:2021vg} find localized regions of high normal stress that propagate at a steady speed in the flow direction. In our previous BSM measurements \cite{Rathee:2017aa,Rathee:2020aa,Rathee:2020un}, we  found a direct connection between high boundary shear stress and normal stress through localized deformation of the compliant boundary. 
Here, by imaging the particles in the bottom layer of the suspension in parallel with local stress measurements we can now see that the dilatant pressure creates a localized region of direct frictional contact between the particles and the shearing boundary and a substantial transient flow in the vorticity direction.
Taken together, these results demonstrate that localized non-affine and dilatant regions are often, and maybe always, associated with strong shear thickening, although the specific dynamics will depend on the details of the system.

\textbf{Dependence on geometry and boundary conditions:} As described above, propagating dilatant fronts are observed in shear thickening suspensions under a variety of conditions, but with detailed behavior that varies depending on the measurement geometry, boundary conditions, and material.  For example, overfilling impacts the measured critical stress for DST  \cite{Fall:2012vd}, and surface properties impact front propagation speeds \cite{Ovarlez:2020ul}.  Similarly, multiple lines of evidence suggest that, as a consequence of the dilatant pressure, the confinement at the edge of the suspension impacts front dynamics \cite{Fall:2012vd, Maharjan:2021ud,Ovarlez:2020ul}.  The results described here are robust and repeatable, and independent of whether the bottom surface is smooth glass or  PDMS (slightly roughened by the fiduciary beads), but in all cases the top surface is a smooth rheometer tool and the suspension extends beyond the outer edge of the rheometer (Fig. \ref{bulkrheology}A).    In our case, the fact that the free surface is in the plane of the top plate  may explain the appearance of fronts whose motion is coupled to the top plate instead of regions that appear coupled to the bottom.  Also, the spatial extent of the fronts is comparable to the rheometer gap, and the appearance of nearly periodic behavior when multiple fronts are present suggests that the fronts interact over relatively long length scales.  Thus, while the existence of heterogeneity in the coupled fields of concentration, stress, and flow speed appears to be ubiquitous in shear thickening, the details of its manifestation will be sensitive to the specific boundary conditions in all three spatial dimensions.

\textbf{Solvent migration:} To our knowledge the results presented above represent the first direct measurement of dynamic solvent migration in sheared suspensions, although the existence of relative flow can be inferred from measurements of concentration fluctuations \cite{Ovarlez:2020ul} and boundary protrusions \cite{Maharjan:2021ud}, and is a necessary consequence of shear-induced dilatancy \cite{Deboeuf:2009uv}.  Liquid migration plays an important role in controlling extrusion of shear thickening suspensions \cite{ONeill:2019tv} as well as in instabilities observed in dense suspensions flowing through constrictions \cite{Genovese:2011wa} or in straight channels \cite{Yaras:1994un, Isa:2009vz,Kanehl:2017vc}.  In particular, measurements on concentrated suspensions flowing in  microchannels found regimes of stable fronts that propagate upstream, with densified regions on the upstream side and rarefied regions behind \cite{Isa:2009vz}, and an associated region of convergent flow that suggests a localized jammed region similar to what we speculate is present in the fronts we observe (see below).  Modeling work by Kanehl \& Stark captures much of the  behavior observed in the microchannels \cite{Kanehl:2017vc}, suggesting that similar approaches could elucidate the propagating fronts we have observed under steady shear.

\begin{figure*}
\includegraphics[width=1\textwidth]{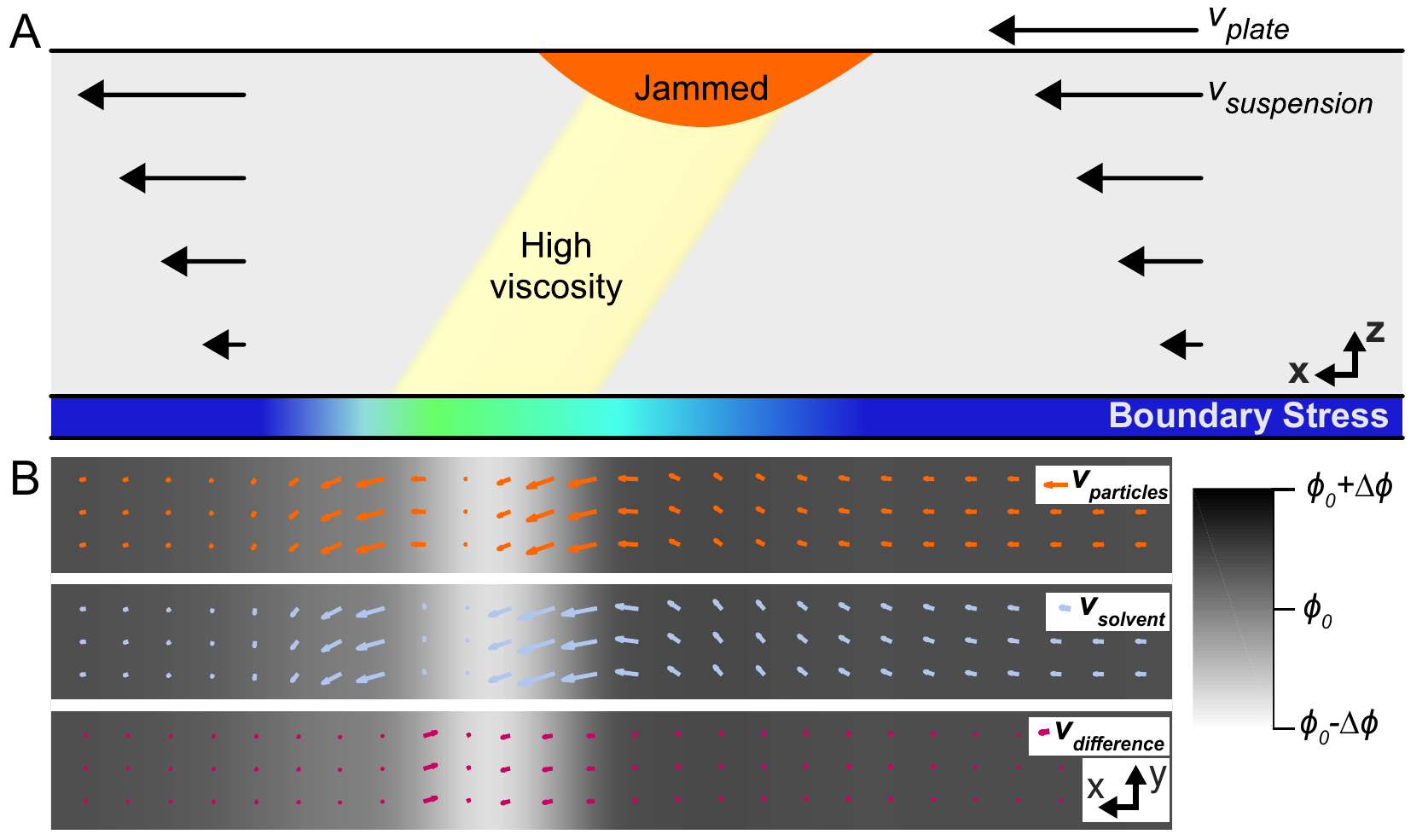}
\caption{ Schematic of model for propagating fronts. (A) Cross-section in the flow-gradient plane.  The suspension flow is uniformly affine away from the front, and the stress at the bottom boundary is low (blue).  The front is anchored by a jammed solid like phase in physical contact with the top boundary, producing a constricted region that results in a transition in the sheared suspension to a high viscosity fluid phase, resulting in high stress at the bottom boundary. (B) Sections of flow fields at the bottom layer of the suspension in the flow-vorticity plane.  The vectors represent the average velocity for the particles (top), solvent (middle), and the difference (bottom), from Fig. \ref{solvent}A and C.  The greyscale represents the relative concentration of the particulate phase inferred from the fluid migration.} 
\label{model}
\end{figure*}

\textbf{Model for propagating fronts:}
The existence of events that have a well defined shape in time determined from the data taken at high magnification, combined with the observation that the high stress regions propagate at a roughly constant velocity when observed at low magnification, allows us to interpret the time profiles shown above as equivalent to spatial profiles, using the propagation speed to convert from time to distance along the flow direction.  
This leads to a possible structure for the propagating fronts, summarized in Fig. \ref{model}.  
We speculate that motion away from the fronts is mostly affine, as indicated by the arrows in the schematic in Fig. \ref{model}A, consistent with the measured uniform low boundary stress and small, constant velocity of the bottom layer of cornstarch particles away from the fronts.  The presence of a stable jammed region or solid-like phase (SLP) in physical contact with the top boundary is inferred from the fact that the fronts propagate with the speed of the top plate, as in Ref \cite{Rathee:2020aa}.  
The SLP generates a region of increased shear rate that pushes the suspension into a high viscosity phase arising from an increase in frictional interparticle interactions, and the resulting high shear stress is reflected in a dramatic increase in stress at the bottom boundary underneath the SLP (the schematic includes the speculation that the region of high stress will be tilted toward the compressive axis, but we have no direct evidence for this).  

Concentration variations accompany the high viscosity region and are associated with variations in the  particle speed, as modeled by the greyscale background in Fig. \ref{model}B, where we assume that concentration changes are simply proportional to the fluid migration.  
Note that this very localized variation is not what would be inferred from the intensity profiles  (Fig. \ref{average}) or the visual observations from the movies that the bottom layer gets gradually denser from the back of one event to the front of another. One possible explanation for this discrepancy is that we only image the bottom layer, which is particularly sensitive to boundary-induced layering, while the fluid flow is presumably responding to concentration variations over a longer length scale, and reflects  particle concentration variation above the bottom layer.

Using the precise measurements of fluid migration (Fig. \ref{solvent}), we can directly estimate the change in pore pressure driving the flow. The extent of the fronts in the flow direction is on the order of the 1 mm gap between the rheometer plates, whether measured directly from the low magnification spatial maps (Fig.   \ref{lowmag}), or inferred as $v_p \Delta t$ from the temporal data measured at high magnification (Fig. \ref{average}), consistent with our previous observations of stress heterogeneities in other shear thickening suspensions \cite{Rathee:2017aa,Rathee:2020aa,Rathee:2020un}.  We can  use this length scale $L\approx1$ mm, in combination with the magnitude of the measured speed of the fluid migration $v_s \sim 100$  $\mu m/s$  to estimate the change in pore pressure  (pressure in the fluid phase) $\Delta P $ driving the flow:  $\Delta P \approx \eta_s v_s L/0.5 a^2$, where $\eta_s$ is the solvent viscosity and $a$ is the average radius of the particles \cite{Maharjan:2021ud,ONeill:2019tv,Isa:2009vz}.  The solvent viscosity is $\eta _s \sim 1\times 10^{-2}$ Pa$\cdot$s, and the average radius of the (highly polydisperse) constarch grains is $a\sim 10$ $\mu$m, so   $\Delta P \sim 20$ Pa. Interestingly, this is comparable to $\sigma _c$, the onset stress for shear thickening (Fig.\ref{bulkrheology}B), which in the Wyart-Cates model \cite{Wyart:2014wy,Wyart.2014zf} is the stress necessary to initiate interparticle frictional contacts and trigger the transition to the high viscosity fluid phase.  The shear and normal stresses in the high viscosity phase are likely more than an order of magnitude larger, and it is not clear why  $\sigma _c$ would control the pore pressure difference (if in fact it does).  Note that the values that enter into the estimate of the pore pressure, the spatial extent of the fronts (Fig. \ref{lowmag}) and the speed of fluid migration (Fig. \ref{solvent}D) are roughly independent of the applied stress or average shear rate.

The striking M-shape of the non-affine velocities (Fig. \ref{average}) is perhaps easier to explain. The initial increase in particle speed may be a direct consequence of the higher shear rate required as the suspension passes through the constriction produced by the SLP (as mentioned above, far from the fronts, the boundary layer velocity is $\sim \dot\gamma R$). That speed drop in the middle of the event arises because the large dilatant pressure produced in the center of the front:  the particulate phase is forced into direct contact with the bottom boundary, so the relative velocity necessarily goes to zero.  This suggests that the central part of the high viscosity region sketched in Fig. \ref{model} includes a region that is transiently jammed, forming a percolating network of high stress that propagates with the top plate. Once the dilatant pressure is released, the particles separate from the boundary and the velocity increases again, reflecting the locally relatively high shear rate and producing the second half of the M-shape.
While the variation in boundary velocity is dramatic in the lab frame, it is important to remember that all of the speeds are small compared to the speed of the top plate, so in the frame co-moving with the top-plate, where all of the fields are approximately time-independent,  particles at the bottom layer are approaching the front at a speed nearly equal to $v_p$ moving from left to right.  
As discussed above, the motion of the particles in  the vorticity direction is consistent with a transient movement of particles towards the edge of the rheometer, suggesting that the front is associated with a bulging out of the particulate phase, consistent with prior observations, as described above.  Note that in the co-moving frame the velocity component in the vorticity direction will be the same as in the lab frame, while the flow component is much larger in magnitude, so in the co-moving frame the change in flow direction will be very slight.  Note also that the relative flow between the particles and the solvent has a small component in the vorticity direction, suggesting that the concentration variation sketched in Fig. \ref{model} gets smaller towards the outer edge of the rheometer.

The proliferation of propagating fronts is responsible for the viscosity increase observed in the bulk rheology in the system studied here. This is reminiscent of our observations of suspensions of colloidal silica spheres \cite{Rathee:2017aa}, where we showed that continuous shear thickening arises from the increasingly frequent presence of propagating localized regions of a high viscosity phase, with a viscosity that increases rapidly with concentration.  In that system, however, we did not observe stable fronts, and the average propagation speed was found to be approximately 1/2 the speed of the top plate.  The remarkable stability of the propagating fronts observed in the cornstarch suspension likely arises when the lifetime of the fronts approaches the rotation period of the rheometer, possibly combined with a stabilizing effect from nearby fronts.

\textbf{Conclusion:} 
We have found a regime in cornstarch suspensions where shear thickening is driven by the appearance of propagating fronts that move at the speed of the top boundary and are associated with large local boundary stress and substantial non-affine flow of particles at the boundary of the suspension.  In addition, we directly measure the relative flow between the particle phase and the suspending fluid and conclude that the fronts contain a localized region of high dilatant pressure, producing a low particle concentration and a pore pressure difference that is on the order of the critical shear stress for shear thickening.  These observations suggest a novel combination of a stable jammed region in contact with one boundary of the system and a propagating front of percolated frictional contacts that spans the gap between the rheometer plates, providing a striking example of the complex structures associated with shear thickening.

\section{Acknowledgements} The authors thank Eric Brown, Annie Colin, Emanuela Del Gado, Benjamin Dolata, Ana\"{i}s Gauthier, and Peter Olmsted for helpful discussions. This work was supported by the National Science Foundation (NSF) under Grant No. DMR-1809890. J. S. U. is supported, in part, by the Georgetown Interdisciplinary Chair in Science Fund.

\section{Methods}
Suspensions were composed of cornstarch particles  (Sigma-Aldrich) suspended in a density matched glycerol-water-cesium chloride mixture (CsCl), with weight fractions calculated from the mass of the constituents. The cornstarch particles are highly polydisperse, with sizes varying roughly from 1  to 50 $\mu$m with a mean size of $\sim$ 20 $\mu$m.   To image the cornstarch particles, 0.02 M fluorescein sodium salt was added to the solution and subsequently absorbed on the particles.  Rheological measurements were
performed on a stress-controlled rheometer (Anton Paar MCR 301)
mounted on an inverted confocal (Leica SP5) microscope \cite{Dutta:2013aa}
using a plate of diameter 25 mm. The gap between rheometer plates was fixed at 1 mm unless stated otherwise. 
For BSM measurements, elastic films of thickness 50 $\pm 3 \mu$m  were deposited by spin coating PDMS (Sylgard 184; Dow Corning) and a curing
agent  on 40 mm diameter glass cover slides (Fisher Sci) that were cleaned thoroughly by plasma cleaning and rinsing
with ethanol and deionized water \cite{Rathee:2017aa}. The PDMS and curing agent were mixed and degassed
until there were no visible air bubbles.  The elastic modulus (G') of the films was $\sim 10$ kPa (55 kPa for the stiff PDMS used to produce supplemental figure 1).   After deposition of PDMS, the slides
were cured at 85 $^{o}$C for 2 hours. After curing, the PDMS was
functionalized with 3-aminopropyl triethoxysilane (Fisher Sci) using
vapor deposition for $~$40 min. For high resolution BSM measurements, carboxylate-modified
fluorescent spherical beads of radius 0.5 $\mu$m with
excitation/emission at 520/560 nm were attached to the PDMS
surface. Before attaching the beads to functionalized PDMS, the beads
were suspended in a solution containing PBS solution
(Thermo-Fisher). The concentration of beads used was 0.006 $\%$
solids. For measurement at lower magnification (larger field of view), 10 $\mu$m beads were attached on the PDMS,  and a second PDMS film of thickness $~$ 5-7
$\mu$m was added by spin coating to prevent bead detachment while shearing.  
Deformation fields were determined with particle image velocimetry (PIV) in ImageJ. 
The surface stresses at the interface were calculated using an extended traction force technique and codes given in ref. \cite{Style:2014aa}.  Taking the component of the surface stress in the flow (velocity) direction, we obtain the  scalar field $\sigma_{BSM}(\vec{r},t)$, representing the spatiotemporally varying surface stress.  

The overall accuracy of the BSM is limited by uncertainty in the modulus of the PDMS layer \cite{Rathee:2020aa}, and measurement noise arises from the resolution of the PIV imaging and residual vibrations in the imaging system. Normalized 2D spatial autocorrelations $g(\vec{\delta r}) = <\delta\sigma_x(\vec{r},t)\delta\sigma_x(\vec{r}+\vec{\delta r},t)> / <\delta\sigma_x(\vec{r},t)^2>$, averaged over $\vec{r}$ and $t$,  are calculated using the Matlab function  xcorr2.  
Temporal cross-correlations, $g(\vec{\delta r} ,\delta t) = <\delta\sigma_x(\vec{r},t)\delta\sigma_x(\vec{r}+\vec{\delta r},t+\delta t)> / <\delta\sigma_x(\vec{r},t)^2>$ are calculated similarly.  The profile of the spatial autocorrelation and spatiotemporal cross correlations along the velocity direction ($g(x)$, Fig.  \ref{lowmag}B and $g(x,\delta t)$, Fig.  \ref{lowmag}C, respectively) are calculated from $g(\vec{\delta r})$ and $g(\vec{\delta r}, \delta t)$ by setting $y=0$ (using the middle 3 rows of pixels, a strip of width 0.26 mm). 
The average stress in the high stress regions (Fig. \ref{segmented}) was calculated as 
$\sigma_{H}  = \langle\sigma_{x}(\vec{r_{i}})\rangle $,
where the average is taken only over the positions $\vec{r_{i}}$  that satisfy the condition that $\sigma_{x}(\vec{r_{i}}) >$ 100Pa.
The values reported are only weakly dependent on the threshold for the value used here. 
For measurements of solvent flow, 0.5 $\mu$ fluorescent microspheres were added to the glycerol-water-cesium chloride mixture before suspending the cornstarch. Solvent flow experiments were preformed on clean glass without the PDMS gel layer, but otherwise the experimental set-up remained unchanged. To maximize the frame rate for the laser scanning confocal, imaging was limited to a 64x512 pixels. Cornstarch particle speeds were measured using PIV, while particle speeds were measured using particle tracking in MATLAB. The sizes of the observed fluorescent puncta indicate that the spheres aggregated into small clumps, but those clumps appeared to be stable for the duration of the experiments.  Upon cessation of shearing the microsphere aggregates exhibited Brownian motion, indicating that they do not adhere to the much larger cornstarch particles.

\bibliography{propagating_fronts}

\clearpage
\newpage

\begin{center}
	\section*{Supplementary Material}
\end{center}

\begin{figure*}
	\renewcommand{\thefigure}{\Alph{figure}}
	\begin{center}
		\includegraphics[scale=0.5]{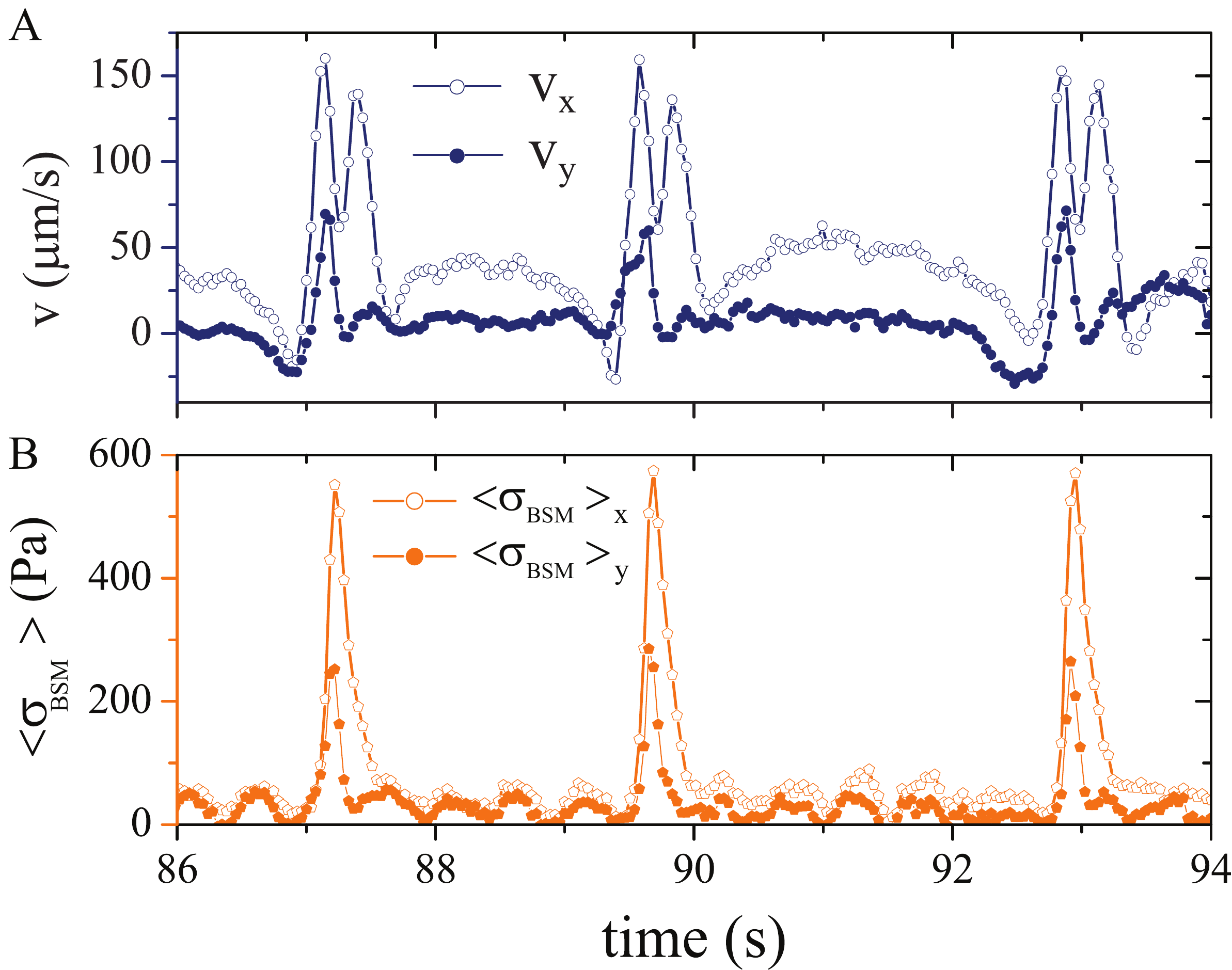}
		\caption{ Cornstarch velocity and boundary stress on stiff PDMS: A)  Average of the $x$ (flow, open circles) and $y$ (vorticity, filled circles) components of the cornstarch velocity determined by PIV of successive images.   B) Components of the boundary stress in the $x$ (open circles) and $y$ (closed circles) directions.   (Stiff PDMS substrate, 100 Pa applied stress).}
	\end{center}
\end{figure*}

\section*{Supplementary Movie Captions}
\noindent\textbf{Supplementary Video 1:} Spatial maps of boundary stress component in the flow direction, $\sigma _x$ at a constant applied shear rate $\dot\gamma=5.5$ s$^{-1}$.

\noindent\textbf{Supplementary Video 2:} Images of the bottom layer of cornstarch particles at 100 Pa applied stress on PDMS coated substrate (first half) and uncoated glass (second half), showing dramatic non-affine motion associated with propagating high stress fronts. 

\noindent\textbf{Supplementary Video 3:} Simultaneous images of the bottom layer of cornstarch particles (left panel) and beads attached to PDMS layer (right panel) at 100 Pa applied stress.  The cornstarch particles are faintly visible in the bead channel, and it is evident that the cornstarch particles move with the PDMS layer for a short period during the events.

\noindent\textbf{Supplementary Video 4:} High speed video of cornstarch and suspended fluorescent microspheres at 120 Pa applied stress showing brief periods of solvent migration during non-affine flow events.  

\end{document}